# Intracavity filtering in SESAM mode-locked fiber lasers: soliton effects and noise performance

JAKUB BOGUSŁAWSKI,[1,*] ŁUKASZ STERCZEWSKI,[1] DOROTA STACHOWIAK,[1] AND GRZEGORZ SOBOŃ[1]

[1]*Laser & Fiber Electronics Group, Faculty of Electronics, Photonics and Microsystems,*
*Wrocław University of Science and Technology, Wybrzeże Wyspiańskiego 27, 50-370 Wrocław, Poland*
*\*jakub.boguslawski@pwr.edu.pl*

**Abstract:** We characterized and analyzed the effect of intracavity spectral filtering in the Er:fiber laser mode-locked with a semiconductor saturable absorber mirror (SESAM). We studied the dispersive properties of bandpass filters and their influence on the characteristics of generated soliton pulses. Our analysis showed that various sideband structures were induced by the filter dispersion profiles and shaped through the interaction of the soliton with the dispersive wave. In addition, intracavity filtering improved the intensity and phase noise of the laser significantly, and we showed optimal filtering conditions for both types of noise. By adding a 10 nm bandpass filter to the laser resonator, the intensity and phase noise were improved 2- and 2.6 times, respectively.

## 1. Introduction

Femtosecond fiber oscillators are a critical part of various systems used in, e.g., frequency comb generation, time distribution, or biomedical applications [1–3]. Increasingly demanding applications require precisely designed laser parameters, such as pulse duration, spectral shape, or noise performance. For the latter, particularly phase noise and timing jitter are essential for frequency metrology [4] or microwave generation [5]. In mode-locked fiber lasers, the source of pulse train timing jitter mainly includes pump-induced gain fluctuations, quantum noise from spontaneous emission in the amplification process, and cavity length fluctuations [6,7]. Strong coupling of pump noise to Kelly sidebands has recently been observed in soliton mode-locked lasers [8]. In addition, using a slow saturable absorber, such as SESAMs, contributes to increased timing jitter [9]. Through the action of a slow saturable absorber, the leading part of the pulse is attenuated more than the trailing part, thus shifting the pulse center backward in each cavity roundtrip. The temporal shift is proportional to the modulation depth of the absorber and the pulse duration [10]. Asymmetrical pulse shaping introduces a temporal shift with magnitude depending on the pulse energy, which provides another mechanism for intensity-to-phase noise coupling, more strongly affecting longer pulses. Despite this disadvantage, SESAMs are still one of the most common solutions used in commercial-grade femtosecond oscillators [11].

Using stable pump sources and all-fiber laser cavities significantly weakens noise levels. Such a cavity can be easily thermally stabilized above room temperature to minimize temperature-related cavity length fluctuations [12]. Even further noise reduction can be obtained by a phase-locked loop to suppress the low-frequency timing jitter. However, active noise suppression adds a layer of complexity. A simple passive strategy for reducing phase noise is dispersion engineering of the cavity. It has been shown that decreasing the cavity net dispersion reduces center frequency noise coupling to timing jitter [13]. However, such an approach requires either bulk dispersive elements [14] or dispersion-compensating fibers [15]. Another difficulty arises from that it requires changing the length of the fibers and, therefore, the pulse repetition rate. An alternative approach to the two above is using intra-cavity programmable filters based on a liquid-crystal-on-silicon platform [16] or a digital micromirror device [17]. Whereas one can obtain dispersion management and suppression of most noisy spectral features, such as Kelly sidebands, devices in this topology are large, complicated, and costly.

A much more straightforward and compact strategy for spectral shaping and noise reduction is intra-cavity filtering with standard filters. A bandwidth-tunable filter has been used to convert amplifier similaritons and dissipative solitons [18]. In a soliton fiber laser working in an anomalous dispersion range, a tunable filter has been used to transition the laser's operation state from single-pulse to double-soliton bound state or double-soliton bound state to triple-soliton bound state [19]. Such an approach also reduces timing jitter from the quantum-limited center frequency noise coupled via dispersion (the Gordon-Haus jitter) [20], and it has been the subject of several works [7,21–23]. In a normal-dispersion nonlinear polarization rotation (NPR) Yb:fiber laser, a significant improvement of noise was obtained using narrow bandpass filtering (i.e., 7-nm bandwidth) [21]. A similar effect has been observed in normal-dispersion Er:fiber lasers, reaching the typical noise level of a stretched-pulse laser [22]. A 2.9-fold reduction of timing jitter has been shown in CNT-mode-locked erbium-doped fiber laser [7]. Since mode-locking with SESAM can introduce additional noise, the effect of intra-cavity filtering has also been studied in such a configuration in Yb:fiber laser, achieving a 5.9 fs (root mean square, RMS) timing jitter in the high-frequency range (> 10 kHz) [23]. However, none of the previous works reports the use of filters to SESAM mode-locked Er:fiber oscillators for noise suppression.

In this paper, we address this niche and show, for the first time, a significant improvement in the noise performance of a SESAM mode-locked Er:fiber femtosecond oscillator via intracavity spectral filtering. We focused on the effect that is usually omitted in this context; according to the Kramers–Kronig relations, the filter's transmission is related to dispersion through a Hilbert transform [24]. A filter might be seen as a compact but highly dispersive element. Such dispersion might affect the spectral and temporal properties of generated pulses. We studied the influence of various filters on the spectral and temporal properties of produced pulses and the interaction with generated dispersive wave components. We showed that this interaction is responsible for forming different sideband structures. As a result, intensity noise was improved 2 times, while phase noise was improved 2.6 times using a 10-nm bandpass filter.

## 2. Experimental setup and methods

Figure 1(a) presents the experimental setup of the investigated laser in an all-fiber and all-polarization maintaining (all-PM) configuration. A 979.5 nm laser diode was used to pump a 56 cm long erbium-doped active fiber (nLight Liekki Er80-8/125-PM) via a 980/1550 wavelength-division multiplexer (WDM). A reflective SESAM (BATOP SAM-1550-33-2ps, modulation depth of 19%, relaxation time of 2 ps) was connected to the cavity using a fiber circulator (CIR), which also enforced unidirectional operation. A 70/30 output coupler extracted 30% of power outside the cavity. We spliced a bandpass filter prior to the output coupler; four configurations were studied: with no filter in the cavity and with 7 nm, 10 nm, and 15 nm filters [see transmittance spectra in Fig. 1(b)–(d)]. The total length of the cavity was about 4.14 m (corresponding to a ~50 MHz repetition rate) and was maintained constant throughout the experiment (see Fig. S1). The net cavity dispersion $D_{2,\text{cav}}$ was anomalous and estimated as −0.087 ps$^2$. To ensure a constant temperature operation, the laser was placed in a 3D printed enclosure (dim. of 15×11×2.5 cm) and on a metal printed circuit board heating plate with uniformly distributed heating traces. The temperature was kept at 28°C. This ensures constant operating conditions.

The laser was next characterized using an optical spectrum analyzer (Yokogawa AQ6380), an optical autocorrelator (APE PulseCheck), frequency-resolved optical gating (Mesa Photonics FROG Scan Ultra2), a radio-frequency (RF) spectrum analyzer (Keysight N9010A) connected with a fast photodiode (Discovery Semiconductors DSC2-50S), and power meter (Thorlabs PD100) with a semiconductor sensor (Thorlabs S122C).

Relative intensity noise (RIN) measurements followed the procedure described in [25,26]. The signals were detected using a photodetector (Thorlabs PDA10CS2) and recorded by an oscilloscope (Rohde & Schwarz RTA4000). The recorded signals were then converted to the frequency domain (within 10 Hz–500 kHz) using fast Fourier transformation and normalized by the average value. We averaged 250 spectra. In the last step, the integrated RIN RMS was calculated by integrating the power spectral density (PSD) over the 10 Hz–500 kHz frequency range.

To measure the phase noise, we employed a signal and spectrum analyzer (Rohde & Schwarz FSW) operating within the spectral range of 3 Hz–1 MHz. The 10$^{th}$ harmonic of the fundamental repetition rate was measured using the InGaAs photodiode connected to the analyzer. Phase noise, $\varphi(t)$, and timing jitter, $x(t)$, are connected through the relation: $x(t) = \varphi(t)/(2\pi v_0)$, where $v_0$ is the carrier frequency [27].

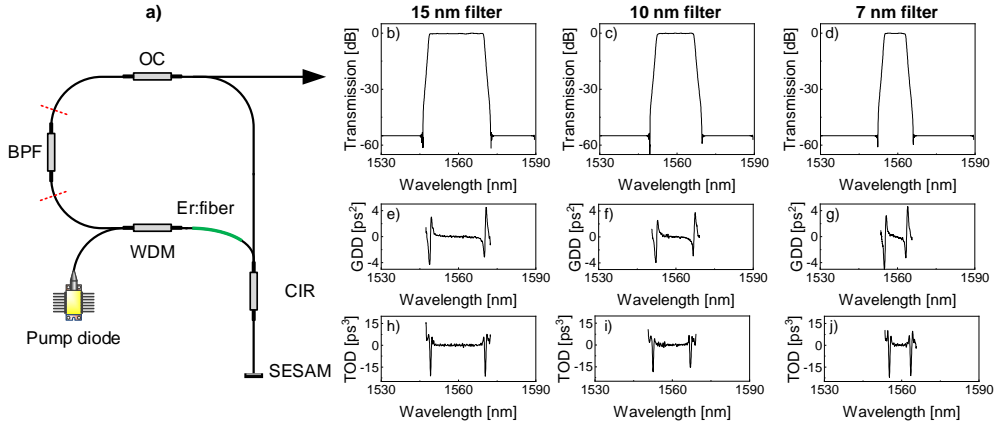

Fig. 1. (a) All-PM mode-locked laser setup. OC – output coupler, CIR – circulator, SESAM – semiconductor saturable absorber mirror, WDM – wavelength-division multiplexer, BPF – bandpass filter. Transmittance spectra of (b) 15nm, (c) 10 nm, and (d) 7 nm filters. GDD profiles of (e) 15 nm, (f) 10 nm, and (g) 7 nm filters. TOD profiles of (h) 15 nm, (i) 10 nm, and (j) 7 nm filters.

The spectral transmittance of fiber filters was characterized using a tunable laser diode source (ANDO AQ4320D) synchronized with an optical spectrum analyzer (ANDO AQ6317B). The group delay dispersion (GDD) and third-order dispersion (TOD) introduced by the filters were numerically calculated using the Kramers–Kronig relations. For convenience, the exact mathematical procedure will be laid out here, as it is central to our findings. Given a measured filter transmittance $T(\omega)$ as a function of optical angular frequency $\omega=2\pi c/\lambda$, where $c$ is the speed of light and $\lambda$ is the vacuum wavelength, one first converts it into the imaginary susceptibility through

$$\chi_I(\omega) = \frac{-c \cdot \log(T(\omega))}{\omega L}. \tag{1}$$

$L$ in the equation is the filter length assumed to be unitary ($L=1$) [28]. Next, one numerically computes a Hilbert transform of $\chi_I(\omega)$ to obtain the real part of the susceptibility $\chi_R(\omega)$

$$\chi_R(\omega) = \hat{\chi}_I(\omega) = \frac{1}{\pi} \int_{-\infty}^{\infty} \frac{\chi_R(\omega')}{\omega - \omega'} d\omega' . \qquad (2)$$

Here, we rely on the Fast Fourier transform (FFT) implementation of the Hilbert transform. The refractive index can then be conveniently retrieved from $\chi_R(\omega)$ through

$$n(\omega) = \sqrt{\chi_R(\omega) + 1} . \qquad (3)$$

In the context of optical pulse propagation, of relevance are the GDD, TOD, and higher-order dispersion terms rather than the spectral phase $\varphi$, which follows the equation

$$\varphi(\omega) = \frac{(n-1)\omega L}{c} . \qquad (4)$$

The frequency-dependent GDD is calculated as the second derivative of $\varphi$ with respect to $\omega$

$$D_2(\omega) = \frac{\partial^2 \varphi}{\partial \omega^2} , \qquad (5)$$

while the TOD is

$$D_3(\omega) = \frac{\partial^3 \varphi}{\partial \omega^3} . \qquad (6)$$

These profiles are later used to establish intracavity phase relationships governing the spectral shape. Figure 1(b)-(j) illustrates the measured filters' transmittance spectra and calculated GDD and TOD profiles for each filter. Each filter introduces strong modulations in GDD and TOD close to the edges of transmittance.

## 3. Results and discussion

### 3.1 Effect on the optical spectrum

Here, we propose a mechanism in which spectral and temporal properties of soliton pulses are manipulated in two ways simultaneously. The first mechanism is direct amplitude modulation resulting from the filter transmittance profile $T(\omega)$. The second mechanism is indirect, through the interaction of the soliton pulse with the dispersive wave.

The dispersion profile of a filter introduces certain phase relationships between the soliton and dispersive wave. Soliton has a flat phase profile, as the cavity dispersion is balanced by nonlinearity:

$$\varphi_{soliton} = -\frac{\bar{D}_2}{2} \tau_0^{-2} , \qquad (7)$$

which depends on the average GDD ($\bar{D}_2$, here retrieved from a fit to the total cavity phase in the region where the soliton propagates), and on the experimentally measured pulse width $\tau_{FWHM}$ [29] such that $\tau_0 = \tau_{FWHM} / (2\log(1+\sqrt{(2)}))$.

On the contrary, the dispersive wave propagates linearly due to much lower intensity and acquires a spectral phase imparted by the cavity dispersion, including that of the filter. Here we obtain it by double-integrating frequency-dependent $D_2(\omega)$ around the soliton center frequency $\omega_0$ with superimposed filter-free net cavity dispersion $D_{2,cav}$. As a result, we obtain the spectral cavity phase relative to the center frequency

$$\varphi_{cavity}(\omega - \omega_0) = \iint \left( D_{2,cav} + D_2(\omega' - \omega_0) \right) d(\omega')^2 . \qquad (8)$$

Note that because of double differentiation in the calculation of $D_2(\omega)$, $\varphi_{cavity}$ is free of a constant offset or linear phase trend (due to group delay) that would be irrelevant to pulse propagation if one directly plugged $\varphi(\omega)$. This phase was next compared with $\varphi_{soliton}$.

Depending on the dispersion profile of the filter, different resonant structures might be formed in the spectrum. A peak-type sideband is formed when the phase difference between the soliton and dispersive waves is a multiple of $2\pi$, resulting in constructive interference. When the phase difference between soliton and dispersive waves is $(2m + 1)\pi$, where $m$ is an integer ($m$ = 0, 1, 2, ...), local destructive interference will occur, resulting in the formation of dip-type sidebands on spectra [30].

We first consider the effect of spectral filtering on the optical spectrum and pulse duration; all configurations are compared at a similar level of estimated intracavity pulse energy at the output coupler (mean = 237 pJ, SD = 7 pJ). Figure 2 and Table 1 summarize the parameters of generated pulses in all configurations, and the top two rows of the figure illustrate the GDD of the filter and the total cavity phase for the studied configurations (while Fig. S2 compares the cavity phase without a filter with filter's phase and soliton phase). Grey dashed lines relate the position of the sideband with respect to the filters' transmittance and phase profile of the cavity. Spectral filtering induces amplitude modulation that also induces chromatic dispersion, as shown by the Kramers–Kronig relations and Fig. 2(a)–(c). We find that the filter-induced GDD significantly contributes to the combined cavity dispersion.

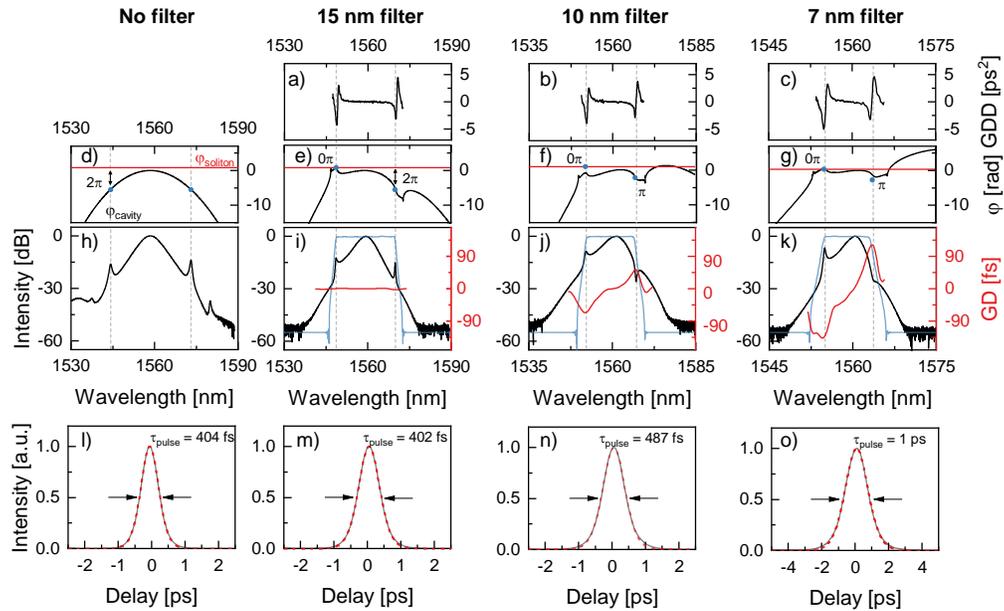

Fig. 2. Intracavity filtering influences spectral and temporal properties of generated ultrashort pulses. Calculated GDD profile of the filter [(a)-(c)], calculated total cavity phase ($\varphi_{cavity}$) and soliton phase [$\varphi_{soliton}$, (d)-(g)], optical spectrum, filter's transmittance (blue curve), and calculated group delay of the pulse [(h)-(k)], and autocorrelation trace [(l)-(o)] of output pulse generated without a filter, and with introducing a 15 nm, 10 nm, and 7 nm filters in the cavity. Gray dashed lines indicate the position of the sidebands with the transmission spectrum of the filter, its GDD, and the total phase of the cavity. Blue dots highlight certain phase relationships between the soliton and the dispersive wave.

When no filter was present, the cavity had a parabolic phase profile ($D_{2,cav}$ = −0.087 ps$^2$, higher-order dispersion terms are neglected). Figure 2(h) and (l) shows the optical spectrum and corresponding autocorrelation of the generated pulses without spectral filtering. The laser produced typical, transform-limited soliton pulses with symmetric Kelly sidebands. Those resonant sidebands are formed where the phase difference between the soliton and the cavity (hence the dispersive wave) equals $2\pi$. Introducing the 15 nm filter modified the total cavity phase [Fig. 2(e)] but did not affect the full-width at half-maximum (FWHM) bandwidth [Fig. 2(i)] and pulse duration [Fig. 2(m)]. The spectral phase profile of the output pulse was flat, as expected for a transform-limited soliton pulse [see also Fig. S3(a) and (d)]. However, we note that the first-order sidebands shifted to fit within the transmission range of the filter, and they colocalize with strong dips in the filter's GDD profile. This potentially suggests that

TOD [31] and higher-order dispersion terms [32] play a role in the formation of sidebands by acting as a strong perturbation of the soliton. The left sideband is formed at 0π, while the right sideband is formed at 2π phase difference and is much sharper. Although the filter reduced amplified spontaneous emission (at 1530 nm), it also cut higher-order sidebands so that only two quasi-symmetrically-located ones could be observed.

Narrower filtering, i.e., 10 and 7 nm, gradually narrowed the spectrum and increased the pulse duration. This was also reflected in the group delay profile of those pulses [Fig. 2(j) and (k)]. For the last, narrowest filter, the temporal intensity profile of the pulse has visibly deviated from the transform-limited profile [Fig. S3(f)]. In the case of the 10 nm filter, the left sideband was formed close (but not precisely) to 0π phase difference at the dip location in the filter's GDD profile. The right sideband, however, was at π phase difference. This results in destructive interference and the formation of a dip-type sideband. Here, the position of the dip-type sideband colocalizes with the point where the filter's GDD profile changes its sign. In the case of the 7 nm filer, the left sideband was formed at the 0π phase difference, while the right sideband was formed close to π phase difference. The sideband was also less pronounced than the right sideband with a 10 nm filter. Previously, it has been shown that the position of the dip sideband is associated with the pulse duration; a dip sideband is closer to the soliton spectrum's peak as the pulse duration increases [30]. This effect was reflected in our data. Finally, we hypothesize that we can see a dip-type sideband because we introduced soliton perturbation (filtering) after the saturable absorber, close to the laser output. Otherwise, the weak intensity dispersive wave component would not pass the saturable absorber.

It is also important to note that for the considered filters, the GDD is almost antisymmetric with respect to the center wavelength. Within the transmission region, the GDD corresponds to normal and anomalous dispersion at shorter and longer wavelengths, respectively. This potentially explains why we observe a gradual wavelength shift towards longer wavelengths, where the negative GDD supports soliton formation [see Table 1].

**Table 1. Comparison of laser operation parameters for the investigated configurations.**

| | Repetition rate [MHz] | Pump current [mA] | Output power [mW] | Intracavity pulse energy [pJ] | Central wavelength / center of mass [nm] | Bandwidth [nm] | Pulse duration [fs] | TBP |
|---|---|---|---|---|---|---|---|---|
| No filter | 50.772 | 177 | 3.46 | 227 | 1558.5 / 1558.5 | 6.3 | 404 | 0.315 |
| 15 nm filter | 50.296 | 182 | 3.55 | 235 | 1560.9 / 1558.9 | 6.4 | 402 | 0.315 |
| 10 nm filter | 50.056 | 196 | 3.63 | 242 | 1560.9 / 1560.2 | 5.3 | 487 | 0.315 |
| 7 nm filter | 50.324 | 191 | 3.67 | 243 | 1560.3 / 1559.8 | 2.5 | 1040 | 0.320 |

We have calculated the GDD and TOD profiles for 3 synthetic filter profiles provided in Fig. S4–S6, and we always observe dispersion oscillations with alternating signs with more gradual changes for sharper-edged filters.

*3.2 Effect of spectral filtering on noise properties*

Now, we consider the effect of intracavity filtering on the noise performance of the laser. Table 2 summarizes the properties of the investigated configurations. Figure 3 demonstrates the effect on the intensity noise, which is substantially decreased. This was reflected in the integrated value of the RIN; it decreased from 0.0303% without any filtering to 0.0142% for the 10 nm filter, which is the most optimal. Further filtering (i.e., with a 7 nm filter) did not improve the noise properties of this laser, and integrated RIN increased to 0.0224%.

A similar trend was visible in the phase noise measurements, as shown in Fig. 4. Intracavity filtering gradually decreased the phase noise of the laser, and, similarly to the intensity noise case, 10 nm filtering bandwidth was optimal. The integrated RMS timing jitter was reduced from 4.45 ps (without a filter in the cavity) to 1.71 ps (with the 10 nm filter in the cavity). Interestingly, one can notice that the 10 nm filter decreased noise at low offset frequencies (<150 Hz) but increased at higher offset frequencies (>150 Hz) when compared to the case without filtering. Further spectral filtering increases timing jitter to 3.93 ps for the 7 nm filter.

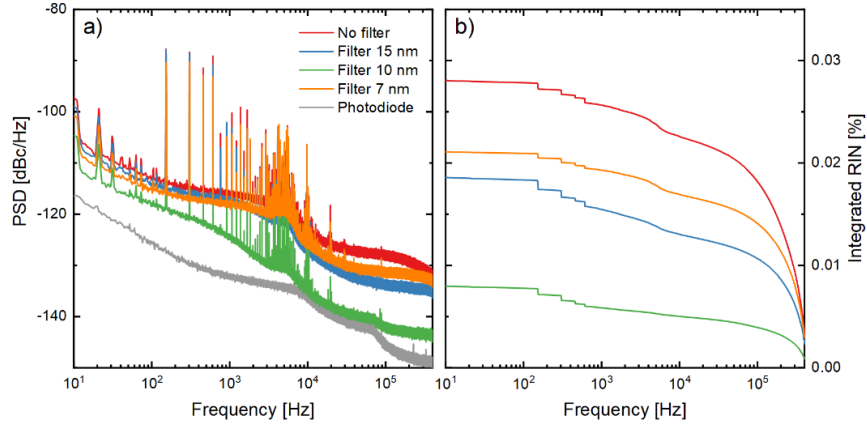

Fig. 3. Characterization of the intensity noise. Power spectral density of the intensity noise (a) and integrated relative intensity noise (b) of the oscillator in all four configurations.

Results show that intracavity spectral filtering substantially improved the noise performance of a free-running mode-locked fiber laser. Intensity and phase noise was improved over 2- and 2.6-fold, respectively. We showed that there are optimal filtering conditions for both the intensity and phase noise performance of the laser.

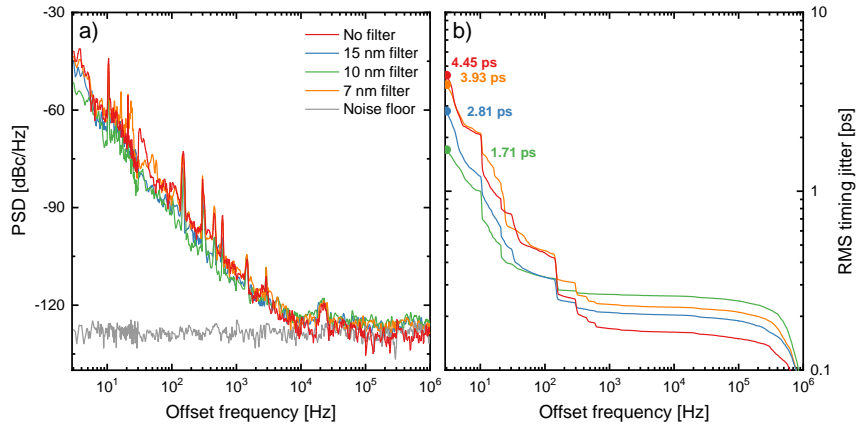

Fig. 4. Characterization of the phase noise. Power spectral density of the phase noise (a) and integrated RMS timing jitter (b) of the oscillator in all four configurations (measured within the 3 Hz – 1 MHz range, at the 10$^{th}$ harmonic of fundamental repetition rate).

We can consider several aspects in which such filtering could improve noise performance. According to the theory developed by Haus and Mecozzi [33], when different spectral shapes of the gain and the loss force the laser to operate away from the maximum of the gain curve, gain fluctuations cause wavelength fluctuations—such a mechanism couples intensity and phase noise through the intracavity dispersion of the laser. In such a case, intensity fluctuations cause central wavelength fluctuations. Due to intracavity dispersion, the latter means different cavity roundtrip times of the pulse, i.e., phase noise. On the one hand, bandpass filtering before the output provides restoring force for mean wavelength by narrowing the gain bandwidth. This decreases timing jitter through the mechanism described above. Besides the classical noise sources, the quantum noise from spontaneous emission is also present. Bandpass filtering helps reduce it outside the spectral range where the soliton pulse is formed.

On the other hand, when the gain bandwidth is further narrowed, the pulse gets stretched in time due to limited spectral bandwidth. Besides gain bandwidth, a diffusion constant of quantum noise is also dependent on pulse duration [7,33], and, at some point, improvement originating from gain narrowing is deteriorated by pulse elongation. Qualitatively similar behavior has been observed previously in a carbon-nanotube mode-locked fiber laser [7]. Optimal filtration conditions are therefore defined by the balance of gain narrowing and temporal stretching – in other words, as much filtration as possible without substantial pulse elongation. Introducing spectral filters also modified the cavity GDD, which might be another effect contributing to the observed noise performance of all four configurations. Particularly, the 7 nm filter significantly increased the cavity GDD at the central wavelength of the pulse (−0.237 ps2 at 1560.3 nm, while Table S1 summarizes these values for all configurations). In this case, the pulse is longer due to limited spectral bandwidth and also slightly chirped (as reflected in the TBP value in Table 1). This might be another effect contributing to the increased noise for the narrowest filter.

Table 2. Comparison of noise properties of the investigated configurations.

|  | Timing jitter 3 Hz – 1 MHz [ps] | Integrated RIN [%] |
|---|---|---|
| No filter | 4.45 | 0.0303 |
| 15 nm filter | 2.81 | 0.0237 |
| 10 nm filter | 1.71 | 0.0142 |
| 7 nm filter | 3.93 | 0.0224 |

## 4. Summary

In this paper, we investigated the effect of spectral and temporal shaping of soliton pulses by intracavity spectral filtering. We demonstrated that intracavity filtering modifies the soliton spectrum in two ways. The first is simple amplitude modulation. The second mechanism is more complex – following amplitude modulation, the filter facilitates dispersive wave generation. Through additional GDD of the filter, various phase relationships between the soliton and the dispersive wave can occur, resulting in different sideband formations. We hypothesize that it might be possible to design the filter properties in a way to shape the pulse profile and spectrum suitably for, e.g., further amplification.

In addition, intracavity spectral filtering is an easy way of significantly improving the noise properties of SESAM mode-locked oscillators. It does not require changing the intracavity dispersion of the cavity by adjusting the lengths of the fiber (that implies changing the repetition rate) or adding active stabilizing elements, such as piezo-stretchers. Intensity noise was improved 2 times, while phase noise was improved 2.6 times. An additional advantage of such a filtering technique is that it stabilizes the central wavelength of the laser, thus improving the reproducibility of laser fabrication.

**Funding.** National Centre for Research and Development, Poland (NCBR, LIDER/32/0119/L-11/19/NCBR/2020), Horizon 2020 Framework Programme (101027721).

**Acknowledgments.** We thank Aleksander Głuszek and Dr. Arkadiusz Hudzikowski for their help in 3D printing the laser enclosure and designing the heating PCB board and electronic driver, and Dr. Jarosław Sotor for helpful discussions. L. A. Sterczewski acknowledges funding from the European Union's Horizon 2020 research and innovation programme under the Marie Skłodowska-Curie grant agreement No 101027721.

**Disclosures.** The authors declare no competing financial interests.

**Data availability.** The MATLAB script generating data from Fig. 1 and Fig. 2 of "Intracavity filtering in SESAM mode-locked fiber lasers: soliton effects and noise performance", and Figs 4-6 from the supplement is available at https://dx.doi.org/10.6084/m9.figshare.25055657. Other data underlying the results presented in this paper are not publicly available at this time but may be obtained from the authors upon request.

**Supplemental document.** See Supplement 1 for supporting content.

# Intracavity filtering in SESAM mode-locked fiber lasers: soliton effects and noise performance: supplemental document

Recorded fundamental beat notes of radio frequency spectrum shows that the repetition rate was maintained similar in all analyzed cases, as shown in Fig. S1. Stable mode-locking operation is confirmed in all cases by a broad spectrum of higher harmonics.

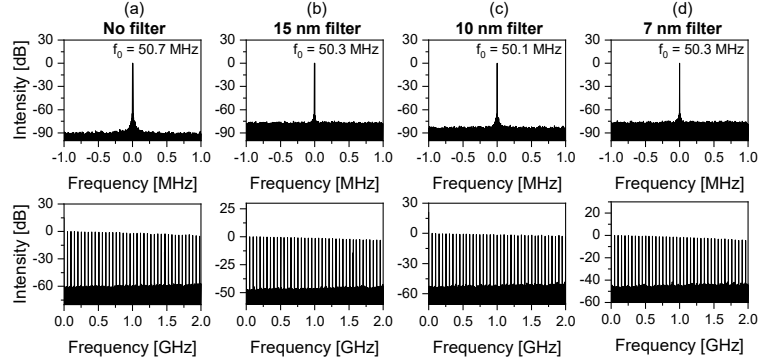

Fig. S1. Fundamental beat note of the radio frequency spectrum (top row) and radio frequency spectrum in a 2 GHz range (bottom row) with (a) no filter in the cavity, (b) 15 nm, (c) 10 nm, and (d) 7 nm filters.

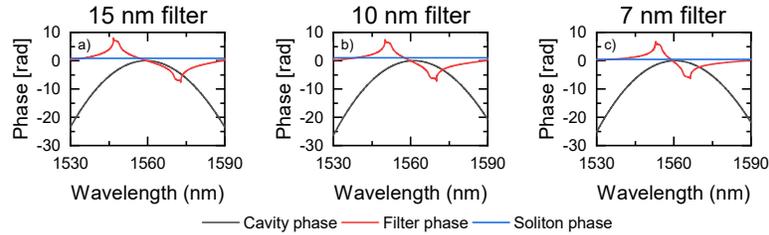

Fig. S2. Calculated total cavity phase (without any additional filter), phase introduced by the filer, and generated soliton for (a) 15 nm filter, (b) 10 nm filter, and (c) 7 nm filter.

Table S1. Comparison of the total cavity GDD (including filter's GDD) at the central wavelength of the pulse for the investigated configurations.

|  | Cavity GDD [$ps^2$] | Central wavelength [nm] |
|---|---|---|
| No filter | −0.087 | 1558.5 |
| 15 nm filter | −0.072 | 1560.9 |
| 10 nm filter | −0.087 | 1560.9 |
| 7 nm filter | −0.237 | 1560.3 |

Figure S3 shows complete temporal and spectral characterization for all three spectral filters. The optical spectrum of the pulse was gradually narrowed, and the pulse acquired a close-to-quadratic spectral phase [Figure S3(a)-(c)]. This is reflected in a progressively elongated pulse, which for the 7 nm filter visibly deviates from transform-limited [Figure S3(d)-(e)]. Measured and retrieved FROG spectrograms are shown in Fig. S3(g)-(l).

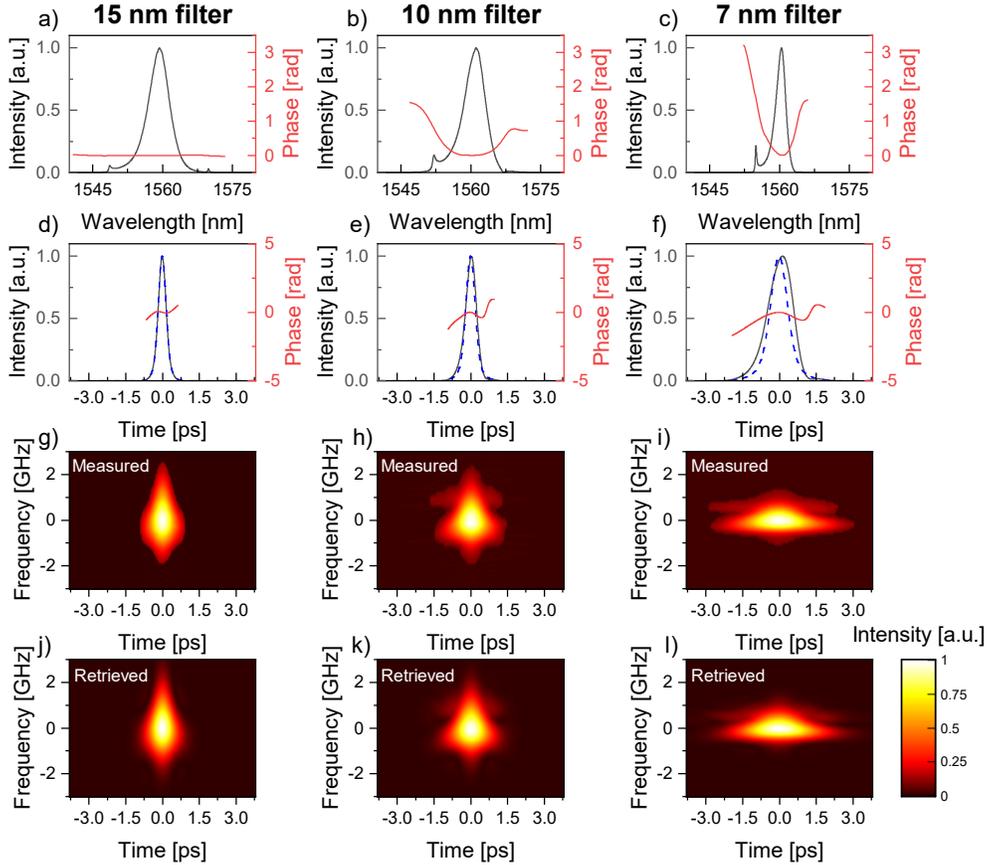

Fig. S3. Spectral and temporal effect of intra-cavity filtration on the generated pulse. Optical spectra (solid gray) and spectral phase profiles (solid red) recorded with 15 nm (a), 10 nm (b), and 7 nm (c) filters. Temporal intensity profile (solid gray), transform-limited intensity profiles (dashed blue), and temporal phase profiles (solid red) reconstructed with 15 nm (d), 10 nm (e), and 7 nm (f) filters. Measured FROG spectrograms of pulses generated with 15 nm (g), 10 nm (h), and 7 nm (i) filters. Retrieved FROG spectrograms of pulses generated with 15 nm (j), 10 nm (k), and 7 nm (l) filters.

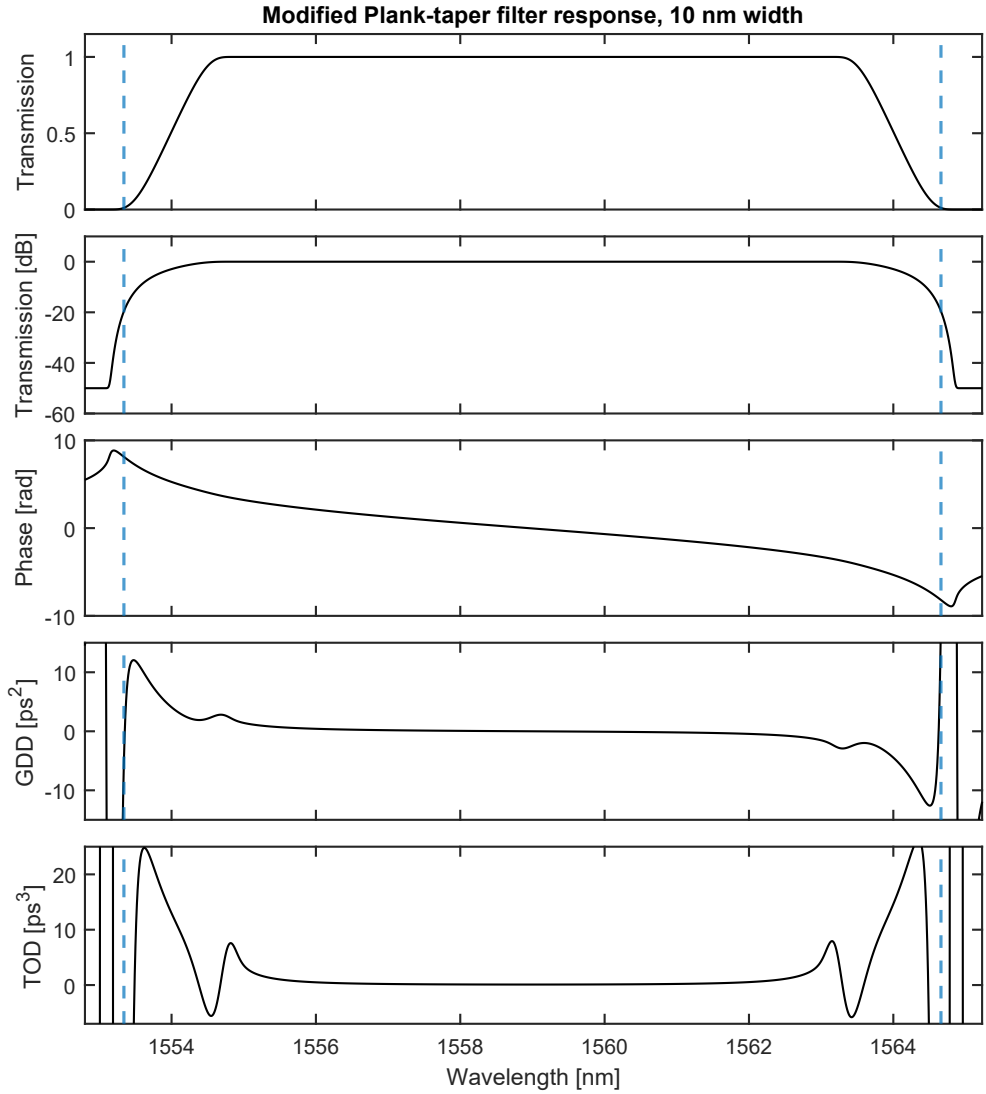

Fig. S4. Synthetic Plank-taper filter response (flat-top with sigmoidal roll-off curves saturating at −50 dB) with a 10 nm full-width at half-maximum (FWHM) and a steepness parameter $\varepsilon=0.1$. The vertical dashed lines denote the −20 dB transmission range. Subsequent panels show the filter phase, group delay dispersion (GDD), and third-order dispersion (TOD). Note the anti-symmetric shape of the GDD with respect to the filter's center wavelength, and the high amount of transmission-induced dispersion due to the knee-like transitions from the passband to the roll-off region.

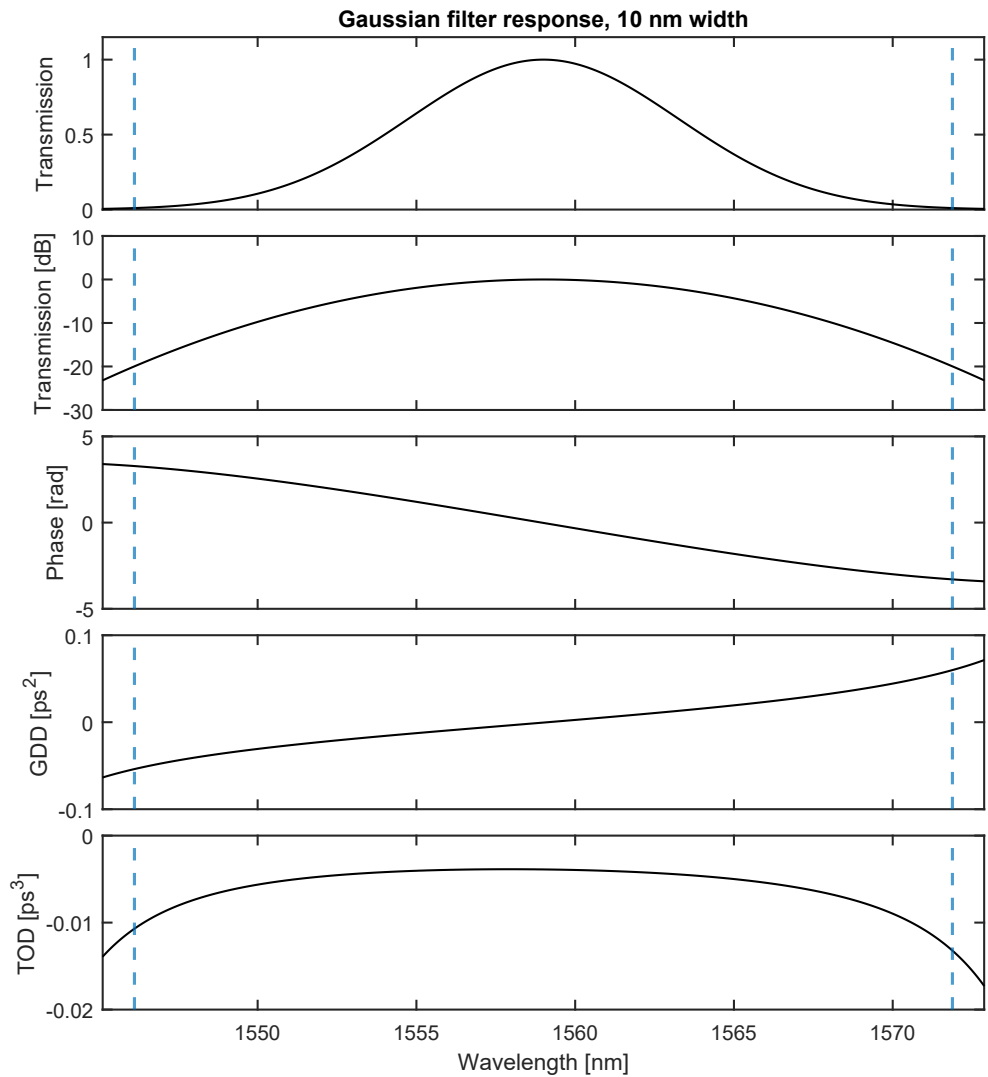

Fig. S5. Synthetic Gaussian filter response with a 10 nm full-width at half-maximum (FWHM) The vertical dashed lines denote the −20 dB transmission range. Subsequent panels show the filter phase, group delay dispersion (GDD), and third-order dispersion (TOD). This filter offers smooth dispersion profiled without rapid oscillations. Notably, the GDD sign is flipped compared to Fig. S3 (rectangular-like Plank-taper window).

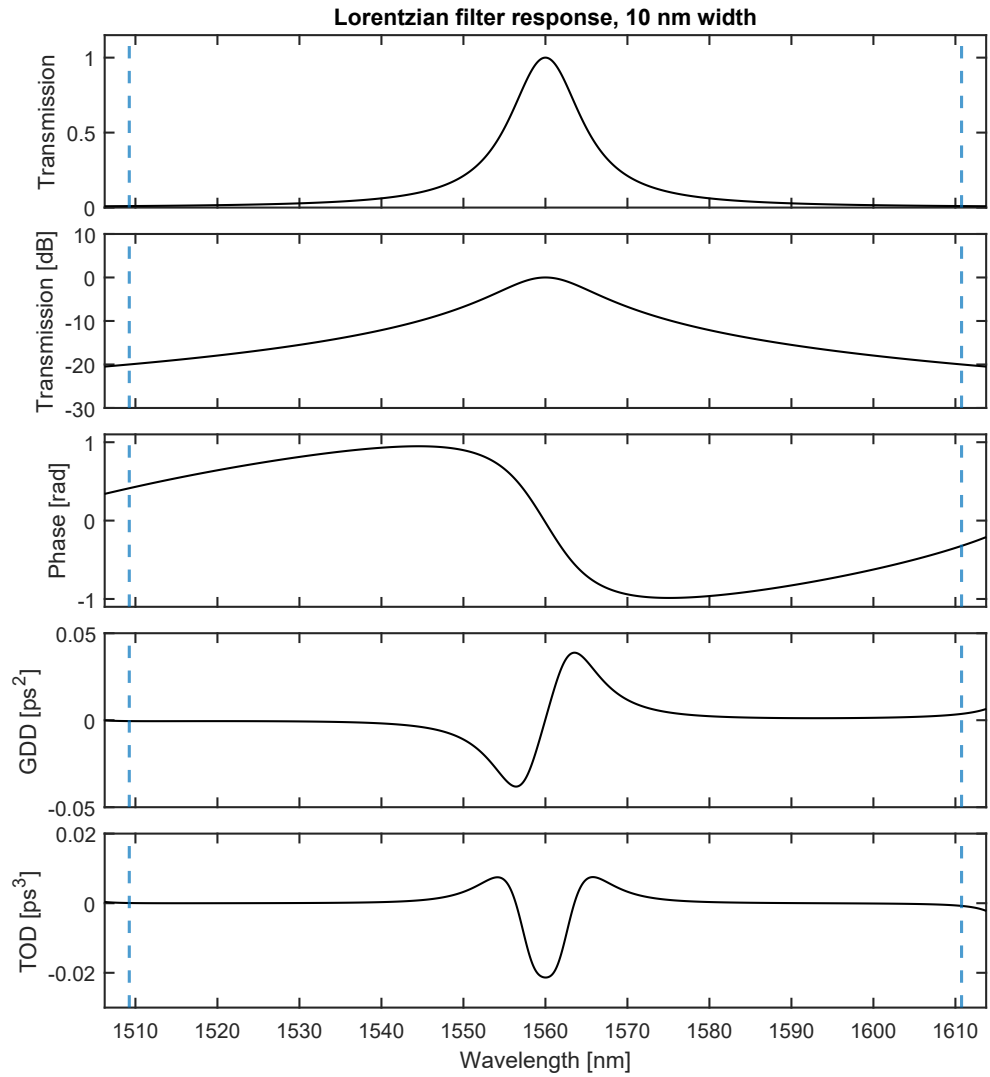

Fig. S6. Synthetic Lorentzian filter response with a 10 nm full-width at half-maximum (FWHM) The vertical dashed lines denote the −20 dB transmission range. Subsequent panels show the filter phase, group delay dispersion (GDD), and third-order dispersion (TOD). This filter offers almost 3×higher maximum GDD in the −3 dB transmission than the Gaussian.